\documentclass[conference]{IEEEtran}
\IEEEoverridecommandlockouts
\usepackage{cite}
\usepackage{amsmath,amssymb,amsfonts}
\usepackage{algorithmic}
\usepackage{graphicx}
\usepackage{textcomp}
\usepackage{xcolor}
\usepackage{hyperref}
\usepackage{subcaption}
\usepackage{float}
\usepackage{multirow}

\def\BibTeX{{\rm B\kern-.05em{\sc i\kern-.025em b}\kern-.08em
    T\kern-.1667em\lower.7ex\hbox{E}\kern-.125emX}}

\DeclareRobustCommand*{\IEEEauthorrefmark}[1]{%
    \raisebox{0pt}[0pt][0pt]{\textsuperscript{\footnotesize\ensuremath{#1}}}}
\begin{document}

\title{FollowUpBot: An LLM-Based Conversational Robot for Automatic Postoperative Follow-up\\
 \thanks{*Equal contribution.}}
\author{\IEEEauthorblockN{Chen Chen\textsuperscript{*}\IEEEauthorrefmark{1},
Jianing Yin\textsuperscript{*}\IEEEauthorrefmark{1},
Jiannong Cao\IEEEauthorrefmark{1},
Zhiyuan Wen\IEEEauthorrefmark{1},
Mingjin Zhang\IEEEauthorrefmark{1},
Weixun Gao\IEEEauthorrefmark{1}, \\
Xiang Wang\IEEEauthorrefmark{1,2},
Haihua Shu\IEEEauthorrefmark{2}}
\IEEEauthorblockA{\IEEEauthorrefmark{1}Department of Computing, The Hong Kong Polytechnic University, Hong Kong, China}
\IEEEauthorblockA{\IEEEauthorrefmark{2}Department of Anesthesiology, Guangdong Provincial People’s Hospital,  Guangzhou, Guangdong, China}
\{chen1.chen, jianing.yin, jiannong.cao, zyuanwen, cs-mingjin.zhang, weixugao, xiang-1988.wang\}@polyu.edu.hk\\
shuhaihua@hotmail.com
}

\maketitle

\begin{abstract}
Postoperative follow-up plays a crucial role in monitoring recovery and identifying complications. However, traditional approaches, typically involving bedside interviews and manual documentation, are time-consuming and labor-intensive. Although existing digital solutions, such as web questionnaires and intelligent automated calls, can alleviate the workload of nurses to a certain extent, they either deliver an inflexible scripted interaction or face private information leakage issues. To address these limitations, this paper introduces FollowUpBot, an LLM-powered edge-deployed robot for postoperative care and monitoring. It allows dynamic planning of optimal routes and uses edge-deployed LLMs to conduct adaptive and face-to-face conversations with patients through multiple interaction modes, ensuring data privacy. Moreover, FollowUpBot is capable of automatically generating structured postoperative follow-up reports for healthcare institutions by analyzing patient interactions during follow-up.  Experimental results demonstrate that our robot achieves high coverage and satisfaction in follow-up interactions, as well as high report generation accuracy across diverse field types. The demonstration video is available at \href{https://www.youtube.com/watch?v=_uFgDO7NoK0}{https://www.youtube.com/watch?v=\_uFgDO7NoK0}.
\end{abstract}


\section{Introduction}

\begin{figure*}[!t]
    \centering
    \includegraphics[width=\textwidth]{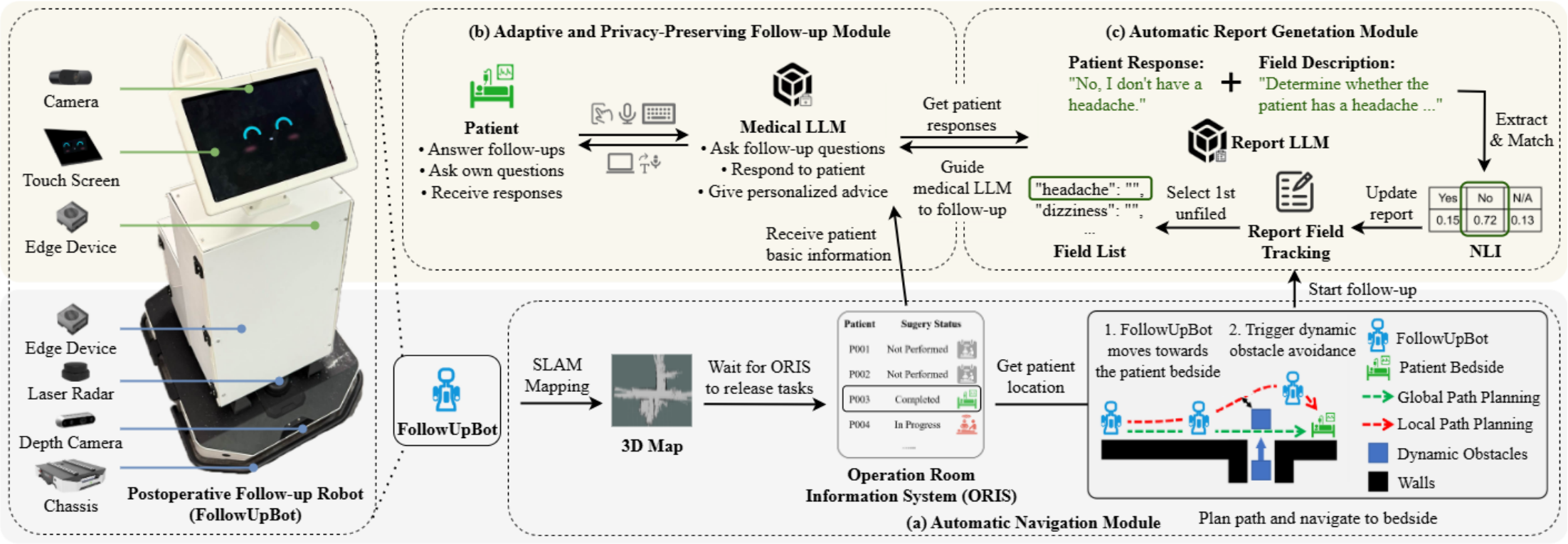}
    \caption{Overview of FollowUpBot, which integrates (a) automatic navigation, (b) adaptive and privacy-preserving follow-up, and (c) automatic report generation into a robotic platform.}
    \label{fig:system-overview}
\end{figure*}

Postoperative follow-up plays a critical role in monitoring patient recovery and identifying potential complications at an early stage \cite{b1}. To achieve this, hospitals typically rely on nurses to perform in-person follow-ups by visiting patients at the bedside and recording their health conditions \cite{b2}. However, the growing number of surgeries and limited clinical staff make the follow-up error-prone and labor-intensive  \cite{b3, b4}. As a result, there is an urgent need for high-quality and labor-efficient approaches to support follow-up.

Despite recent advances in digital follow-up, including automatic platforms, apps and wearables, that ease the burden on nurses \cite{b5, b6, b7, b15}, these approaches still face three major limitations. In particular, they lack physical embodiment, which reduces patient engagement during follow-up interactions and relies on cloud services. This reliance, in turn, raises concerns related to data privacy and clinical compliance \cite{b8, b9}. In addition, many of these platforms are based on scripted interactions that cannot adapt to the specific conditions and emotional responses of patients. Together, these limitations hinder the effectiveness of postoperative follow-up \cite{b10}.

To fill these gaps, we developed an LLM-based conversational robot for automatic postoperative follow-up, named FollowUpBot. The robot is composed of three core modules. \textbf{Automatic Navigation Module} ensures safe and efficient pathfinding to bedside through dynamical obstacle avoidance. \textbf{Adaptive and Privacy-Preserving Follow-up Module} enables adaptive follow-up interaction by generating personalized medical responses based on patients’ clinical profiles and real-time interaction data. To accommodate differences in patients’ postoperative physical and cognitive abilities, the module supports multiple input modalities, including speech, touch, and text. \textbf{Automatic Report Generation Module} supports high-accuracy transformation of dialogue content into standardized postoperative follow-up reports. Specifically, both the follow-up and report generation modules are deployed on edge devices to ensure patient information privacy.

We evaluate FollowUpBot on a synthetic dataset of 100 postoperative cases, focusing on follow-up interaction quality and report generation accuracy. In interactions with GPT-4o simulated patients, the robot achieves 100\% symptom coverage and higher satisfaction scores compared to the WiNGPT2 baseline. Ablation studies show that NLI-based Answer verification and explicit field tracking significantly improve report accuracy and formatting consistency, validating our modular approach.

To summarize, the contributions of this paper are listed:
\begin{enumerate}
    \item As far as we know, FollowUpBot is the first postoperative follow-up robot that integrates navigation, interaction, and report generation modules.
    \item FollowUpBot enables personalized and multimodal follow-up conversations, adapting dynamically to clinical conditions and input preferences of patients. 
    \item FollowUpBot generates accurate, structured follow-up reports from dialogue.
\end{enumerate}

\section{Related Works}

In recent years, digital health technologies have been widely used for postoperative care. Among them, mobile health applications have shown strong capabilities in symptom tracking and medication reminders, effectively supporting nurses and streamlining postoperative follow-up \cite{b5,b12,b13}. Especially during the COVID-19 pandemic, the telemedicine platforms have played an important role in postoperative care, but studies have revealed uneven engagement across demographic groups, particularly among the elderly \cite{b16,b17}.
    
Wearable devices offer a passive, low-effort approach to postoperative monitoring, helping address engagement gaps across patient populations. They enable continuous vital sign tracking and early complication detection \cite{b6,b14}, but in real-world clinical settings, their effectiveness is often limited by concerns about measurement accuracy and long-term patient adherence. AI-based follow-up systems powered by large language models have shown strong potential in postoperative care, enabling high-quality follow-up conversation and comprehensive symptom collection with high patient satisfaction \cite{b7}. However, these AI-driven systems mainly rely on cloud-based APIs \cite{b15}, raising concerns about data privacy, compliance, and latency \cite{b11,b18,b19}. Their clinical deployment is also limited by poor Electronic Health Records (EHR) integration, suboptimal interfaces, and lack of standardization \cite{b20}.

\section{FollowUpBot}

\begin{figure*}[t]
  \centering
  \begin{subfigure}[t]{0.48\textwidth}
    \centering
    \includegraphics[height=3.5cm]{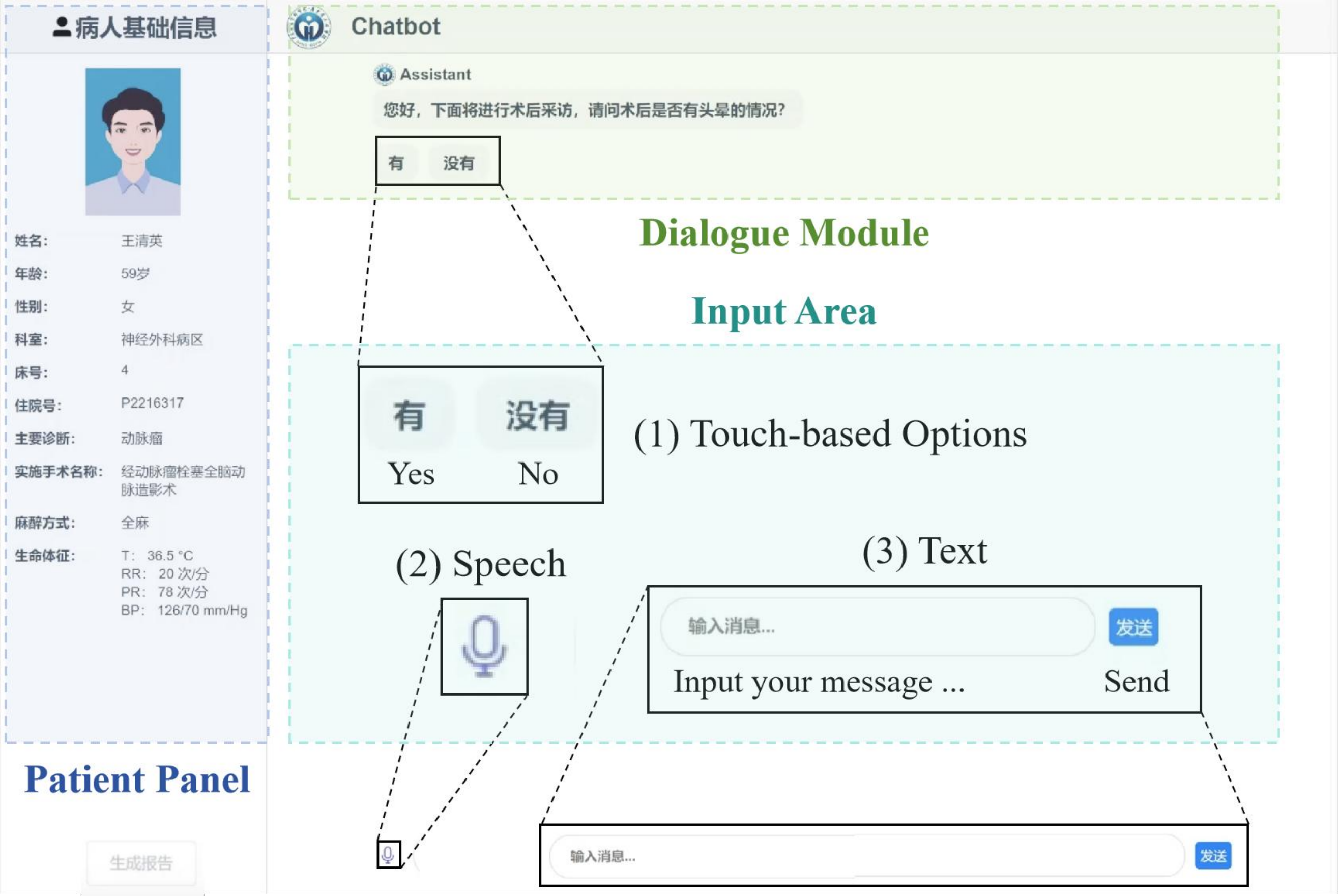}
    \caption{Patient interaction interface for multimodal follow-up dialogue.}
    \label{fig:interface-patient}
  \end{subfigure}
  \begin{subfigure}[t]{0.48\textwidth}
    \centering
    \includegraphics[height=3.5cm]{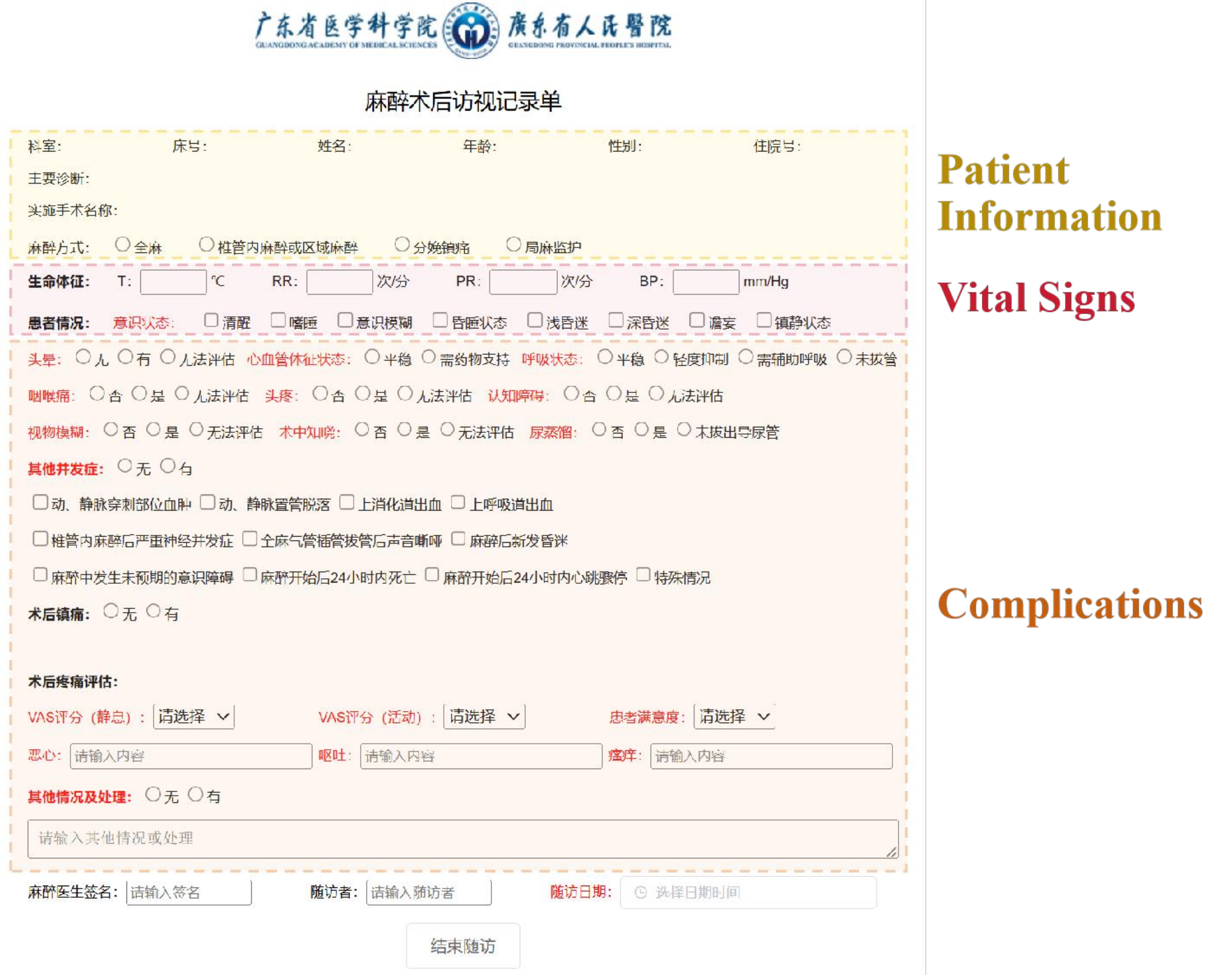}
    \caption{Report generation interface for follow-up results display.}
    \label{fig:interface-report}
  \end{subfigure}
  \caption{Main interactive interfaces of FollowUpBot.}
  \label{fig:system-interface}
\end{figure*}

\begin{figure*}[t]
  \centering
  \captionsetup[subfigure]{justification=centering}
  \begin{subfigure}{0.24\textwidth}
    \centering
    \includegraphics[height=2.9cm]{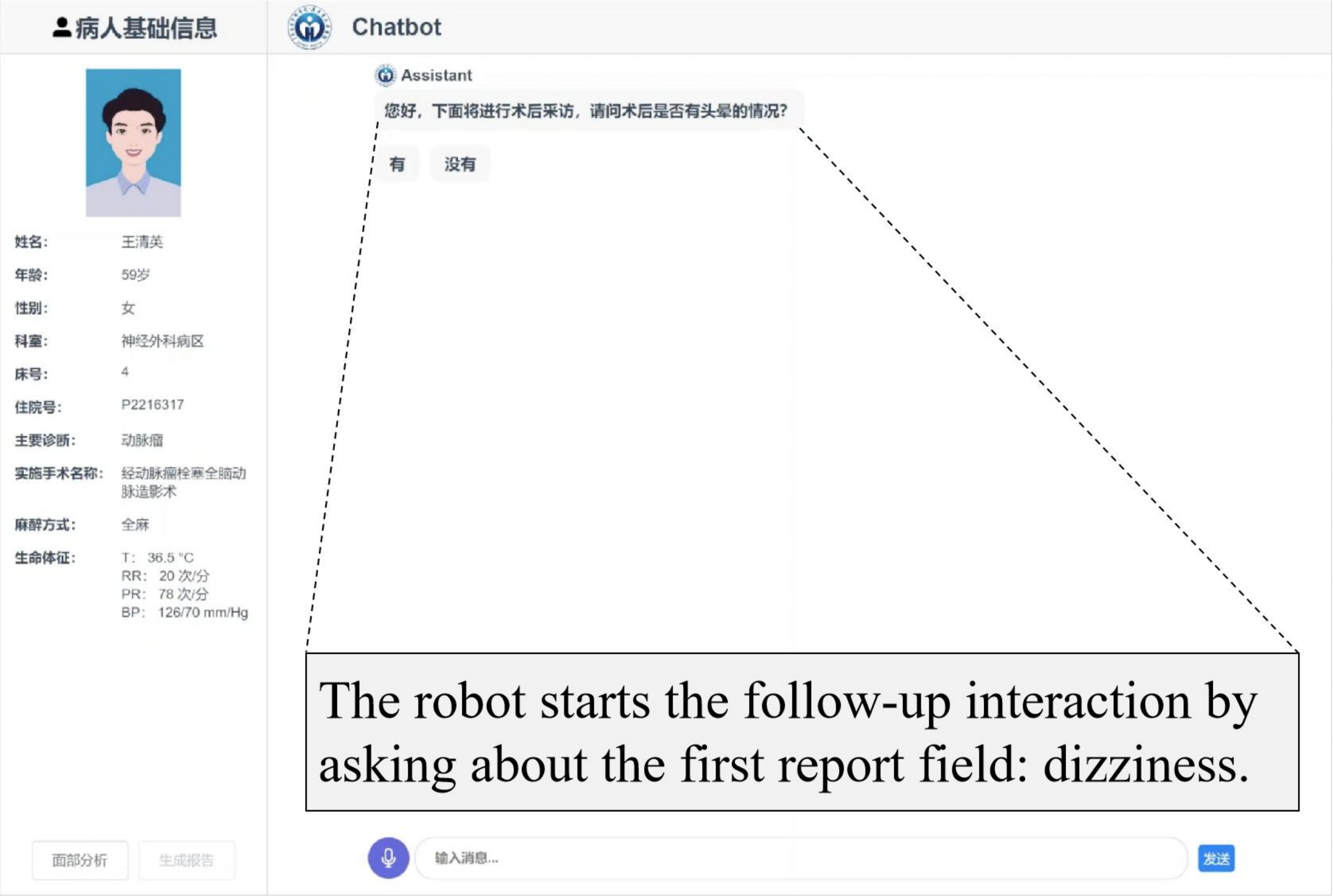}
    \caption{Robot initiates follow-up.}
    \label{fig:demo-a}
  \end{subfigure}
  \begin{subfigure}{0.24\textwidth}
    \centering
    \includegraphics[height=2.9cm]{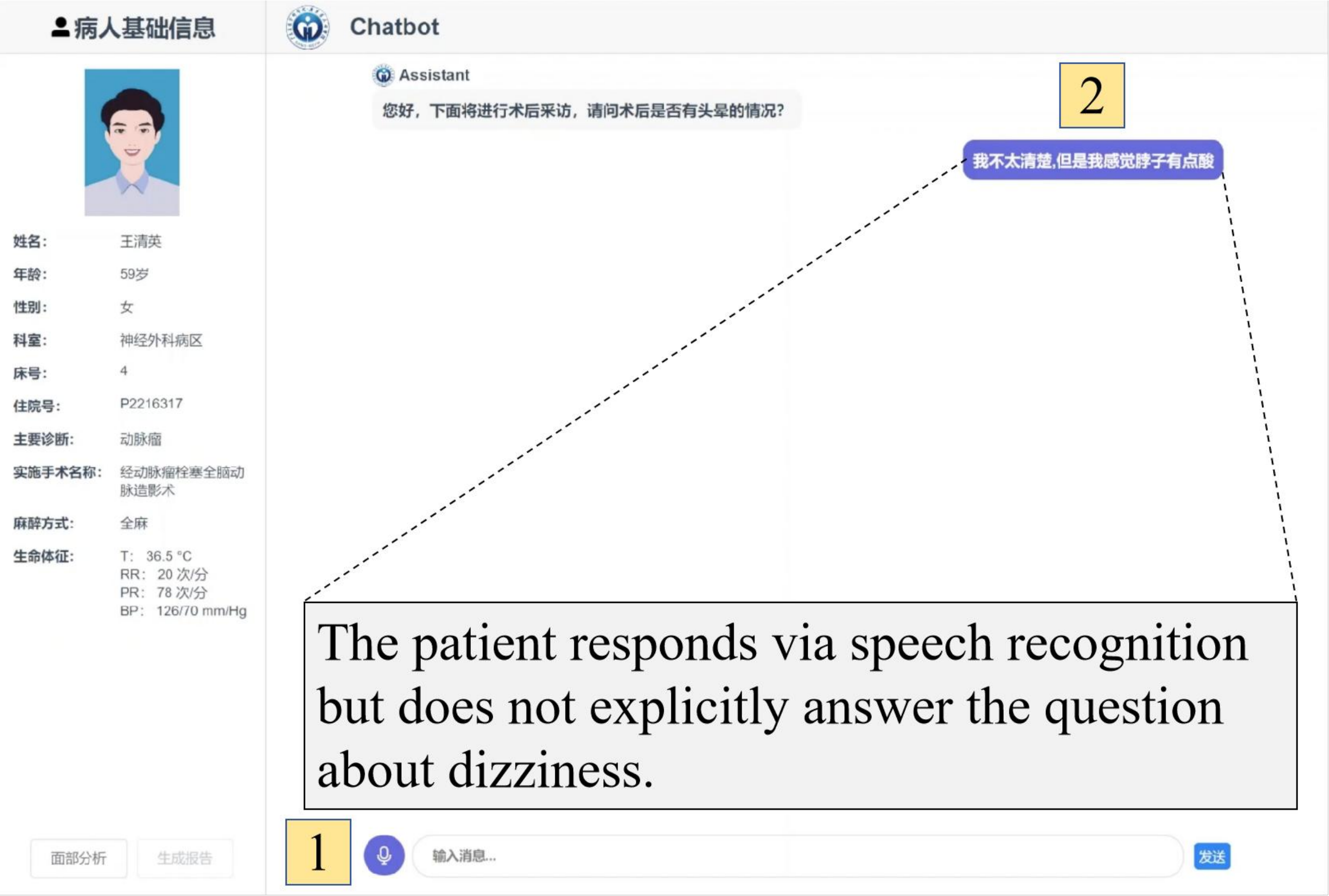}
    \caption{Patient replies via speech.}
    \label{fig:demo-b}
  \end{subfigure}
  \begin{subfigure}{0.24\textwidth}
    \centering
    \includegraphics[height=2.9cm]{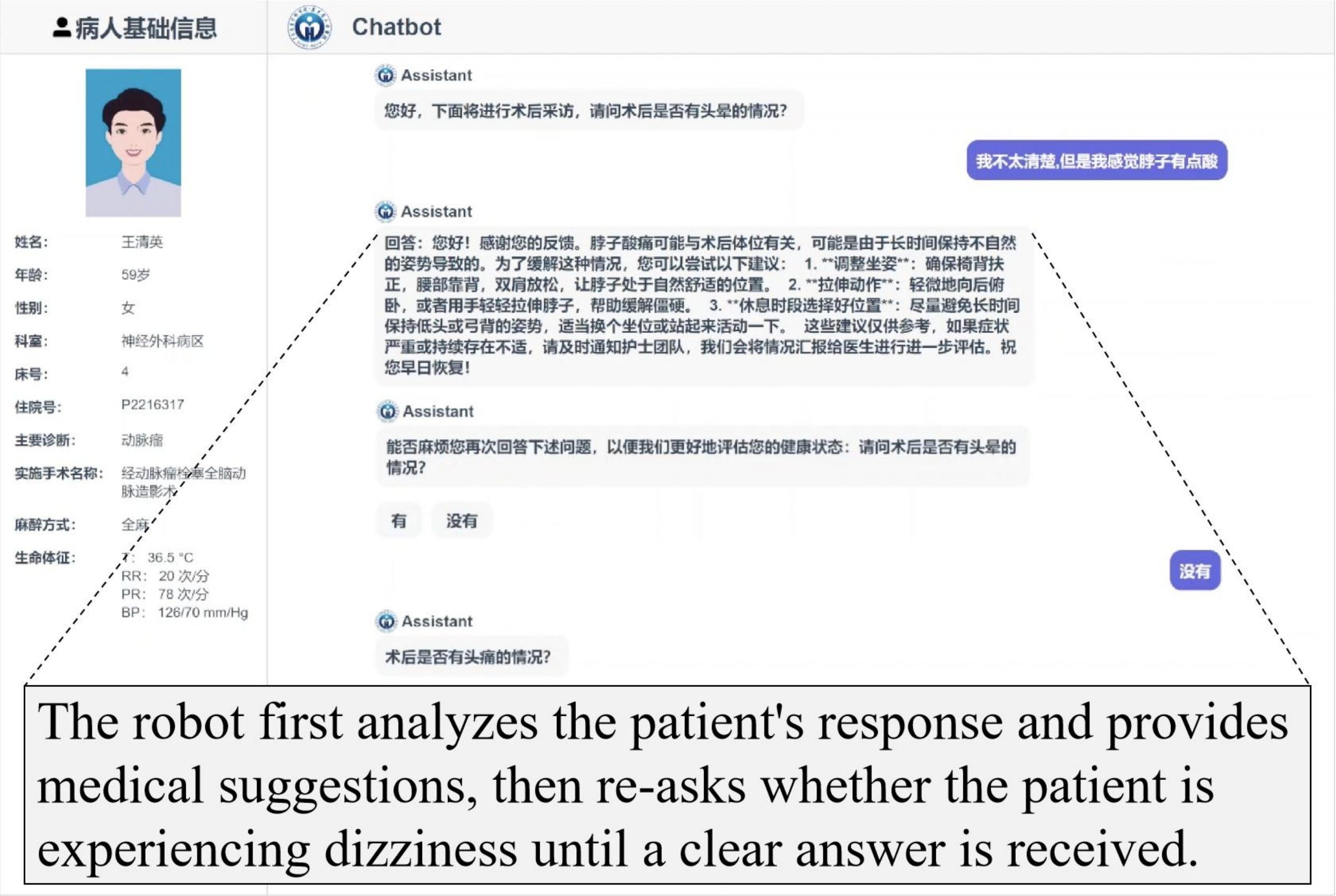}
    \caption{Robot responds to patient.}
    \label{fig:demo-c}
  \end{subfigure}
  \begin{subfigure}{0.24\textwidth}
    \centering
    \includegraphics[height=2.9cm]{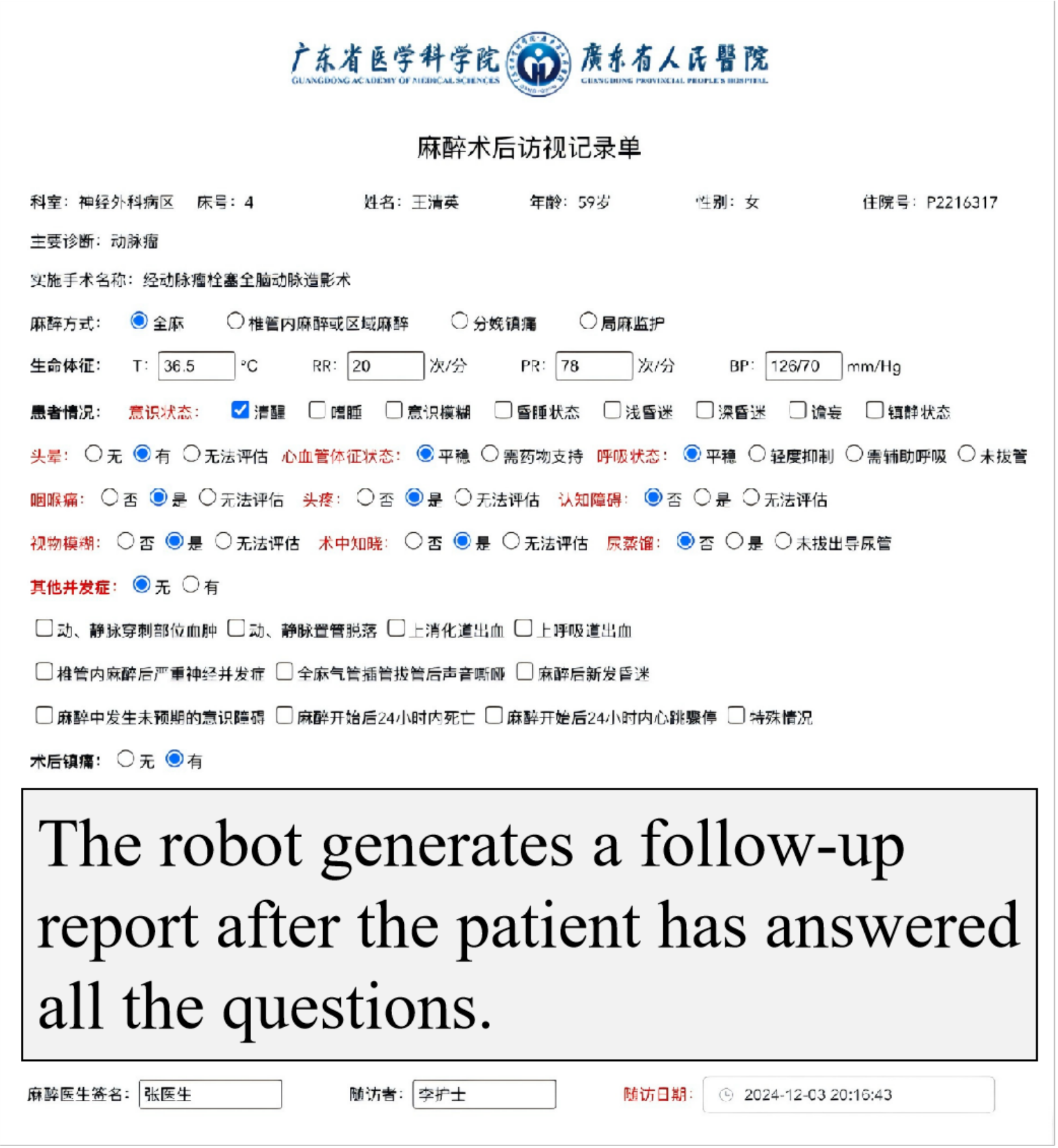}
    \caption{Robot generates report.}
    \label{fig:demo-d}
  \end{subfigure}
  \caption{Execution flow of FollowUpBot. The robot initiates the conversation, receives the patient’s reply, responds, and generates a structured report.}
  \label{fig:demo-interfaces}
\end{figure*}

\subsection{Overview and Workflow}

FollowUpBot is an end-to-end robotic system designed to automate in-hospital postoperative follow-up. As shown in Figure~\ref{fig:system-overview}, the robot consists of three tightly integrated modules: (a) automatic navigation, (b) adaptive and privacy-preserving follow-up, and (c) automatic report generation. The robot is equipped with an RGB camera, touchscreen, audio I/O, multimodal sensors (LiDAR, RGB-D camera), and dual edge devices for local model real-time inference and control.

Upon startup, the robot performs a SLAM-based 3D mapping of the ward and enters a standby state. Once a new follow-up task is issued by the Operation Room Information System (ORIS), the robot retrieves the patient’s ID, bed number, and surgery details. Then it computes a global path to the bedside and dynamically avoids obstacles during navigation.

Upon arrival, the robot initiates a structured dialogue workflow. It maintains a prioritized list of follow-up fields (e.g., headache, dizziness, nausea) and queries each field sequentially using a local medical LLM. The dialogue is dynamically adapted based on patient profile and real-time responses, while patients can interact through speech, touch, or text inputs.

For each field, we provide a detailed description to guide the LLM in extracting appropriate content in the requested format from the dialogue. For fields requiring strict formats (e.g., single-choice fields, numerical fields), a Natural Language Inference (NLI) module is used to map free-form LLM outputs to the closest valid options. Each confirmed field is marked as completed, and the system proceeds to the next.

Once all required fields are filled, a structured report is generated according to the hospital's template and saved locally. The robot then checks for remaining follow-up tasks or returns to standby. This closed-loop workflow enables safe navigation, adaptive interaction, and accurate report generation, which can reduce clinician workload and improve follow-up efficiency.

\subsection{Automatic Navigation Module}

FollowUpBot is deployed on a mobile robot equipped with 2D LiDAR, RGB-D camera, wheel encoders, and an IMU-assisted localization system. It performs online SLAM by fusing LiDAR, RGB-D, and IMU data to build a semantic 3D map of the hospital ward, identifying beds, walls, and restricted zones. Upon receiving a task, it locates the target bed and initiates automatic navigation. A hierarchical planner computes a global path using A*-based search on a topological graph, with costs adjusted by distance and environmental risk. Locally, a Model Predictive Controller (MPC) with a reinforcement learning policy handles trajectory execution and avoids dynamic obstacles. As it approaches the patient, the robot slows down and confirms arrival via LiDAR before handing control to the interaction module.

\subsection{Follow-up Interaction and Report Generation Modules}

\textbf{Multimodal, Privacy-Preserving Follow-up.}
Once at the bedside, the robot initiates a structured follow-up conversation using \texttt{WiNGPT2-Llama-3-8B-Chat}, a locally deployed medical LLM. The model receives the patient's basic information (e.g., age, surgery type) from ORIS to personalize the interaction. Patients can respond via speech (transcribed using \texttt{whisper-large-v3}), touch options, or text input, while the LLM replies are rendered on-screen and via speech synthesis. Each interaction focuses on a specific follow-up field (e.g., “headache”), guided by the dynamic field tracking mechanism.

\textbf{Field Tracking for Structured Completion.}
The robot maintains a list of fields from the hospital’s follow-up template, each with a label, type and description (e.g., “headache”: single-choice, “Determine whether... Yes/No/Unclear”). It dynamically selects the next unfilled field and guides the medical LLM to elicit patient responses. Then, a report LLM (\texttt{Llama-3.1-8B}) extracts field values from dialogue content. For fields with strict formats, we apply an NLI-based postprocessor to verify the output.

\textbf{Answer Verification via NLI.}
We use a cross-encoder NLI model (\texttt{nli-deberta-v3-base}) to map free-form answers to predefined options by computing entailment scores. The option with the highest score is selected as the final report value, ensuring semantic correctness and format consistency.

\textbf{Report Generation and Completion.}
After a field is verified and completed, the process proceeds to the next field. Upon completion of all required fields, the report LLM generates a structured report following the hospital’s format and stores it locally. This modular pipeline, combining on-device LLMs, dynamic field tracking, and template-aware postprocessing, enables accurate, interpretable, and privacy-compliant follow-up automation.

\section{Demonstrations}

In Figure~\ref{fig:system-interface}, we show two main interfaces of the robot: \textbf{Patient Interaction Interface} is designed to support patients multimodal interaction during postoperative follow-up. It includes three key components: a patient panel that displays basic information, a dialogue module that presents follow-up conversations in real time, and an input area designed to support speech, touch, text interactions. \textbf{Report Generation Interface} provides a concise record of postoperative data that integrates patient information, vital signs, and complications into a unified view. To demonstrate how these interfaces function in practice, we present a typical usage scenario of the robot during a postoperative ward round, as shown in Figure~\ref{fig:demo-interfaces}. Please note that this figure focuses on the interaction interfaces only. A comprehensive demonstration of the complete robotic workflow is available in our accompanying video on \href{https://www.youtube.com/watch?v=_uFgDO7NoK0}{https://www.youtube.com/watch?v=\_uFgDO7NoK0}.

\section{Experiment and Evaluation}

FollowUpBot was deployed and tested in Guangdong Provincial People's Hospital, where it successfully navigated real inpatient wards and completed automatic follow-up with real patients, demonstrating its clinical feasibility in real-world hospital environments. 

To quantitatively evaluate robot components, we constructed a synthetic dataset of 100 postoperative follow-up cases using GPT-4o. Each case includes a patient profile, a multi-turn dialogue with field-level annotations, and a structured report.

\subsection{Follow-up Interaction Quality}

\begin{figure}[t]
    \centering

    \includegraphics[height=4cm]{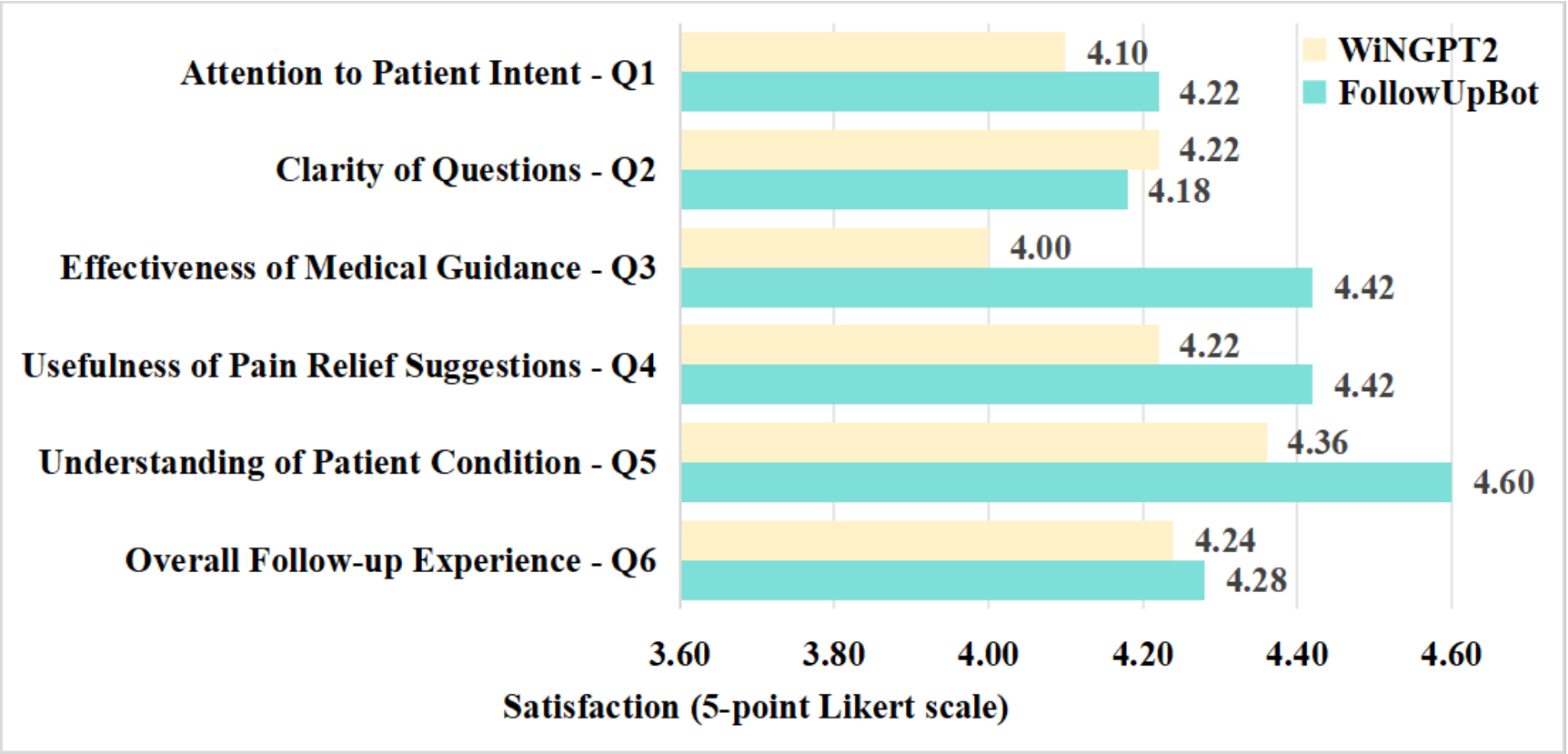}
    \caption{Simulated patient satisfaction across six aspects.}
    \label{fig:satisfaction}
\end{figure}

To evaluate follow-up quality, we simulate patient interactions using GPT-4o based on 100 patient profiles. We compare our robot against a prompting-only baseline (WiNGPT2) across two metrics in follow-up dialogues with simulated patients: coverage and simulated patient satisfaction. Coverage measures the proportion of clinically required symptoms addressed during the dialogue. Our robot achieves 100\% coverage, while WiNGPT2 only covers 53.8\%. Satisfaction is assessed after each dialogue by the simulator across six aspects on a 5-point Likert scale. As shown in Figure~\ref{fig:satisfaction}, our robot outperforms the baseline in most dimensions, confirming both completeness and interaction quality.

\subsection{Report Generation Accuracy}

We conduct ablation studies on three components: field-specific descriptions, NLI-based option matching, and explicit field tracking. Each experiment is repeated five times on a 100-sample dataset, with results averaged. As shown in Table~\ref{tab:report_ablation}, NLI alignment significantly improves accuracy, especially for structured fields. Field tracking further enhances performance, achieving 91.44\% accuracy and 0.9912 BERTScore F1 on single-choice fields, and 99.20\% accuracy with a 0.0300 MAE on numerical fields. Text fields also benefit from more focused dialogue inputs, resulting in moderate gains (0.8512 BERTScore F1). These results highlight the effectiveness of our modular, field-aware report generation pipeline.

\begin{table}[t]
\caption{Report generation performance by field type.}
\label{tab:report_ablation}
\centering
\begin{tabular}{lccc}
\hline
\textbf{Setting} & \textbf{Choice Acc. / F1} & \textbf{Num Acc. / MAE} & \textbf{Text F1} \\
\hline
Desc. Only      & 0.1848 / 0.5852 & 0.5274 / 1.5763 & 0.7290 \\
+NLI            & 0.8216 / 0.9472 & 0.9776 / 0.1506 & 0.7324 \\
Full (Ours)     & \textbf{0.9144 / 0.9912} & \textbf{0.9920 / 0.0300} & \textbf{0.8513} \\
\hline
\end{tabular}
\end{table}

\section{Conclusion}

In this paper, we present \textbf{FollowUpBot}, a robot designed to automatically conduct postoperative follow-up through multimodal interaction and generate structured clinical reports with high accuracy. The robot integrates real-time navigation, medical LLM, field-aware dialogue tracking, NLI-based output normalization, and report generation into a cohesive pipeline. Experimental results show that our robot achieves high coverage and satisfaction in follow-up interaction and improves report generation accuracy across multiple field types.

\end{document}